%% file: dacal_arxiv_v2.0.tex
\documentclass{article} 
\usepackage{iclr2020_conference,times}

\input{math_commands.tex}

\usepackage{hyperref}
\usepackage{url}

\usepackage{times}
\usepackage{epsfig}
\usepackage{graphicx}
\usepackage{amsmath}
\usepackage{amssymb}
\usepackage{caption}
\usepackage{color}
\usepackage{subfig}
\usepackage{lipsum}
\usepackage{bm}
\usepackage{array}
\usepackage{booktabs}

\usepackage{xcolor}

\usepackage{xcolor}

\title{Divide-and-Conquer Adversarial Learning for High-Resolution Image and Video Enhancement}
\author{Zhiwu Huang$^\dagger$, Danda Pani Paudel$^\dagger$, Guanju Li$^\dagger$, Jiqing Wu$^\dagger$, Radu Timofte$^\dagger$, Luc Van Gool$^{\dagger\ddagger}$ \\
	$^\dagger$Computer Vision Lab, ETH Zurich, Switzerland, 
	$^\ddagger$VISICS, KU Leuven, Belgium\\
	\texttt{\{zhiwu.huang, paudel, radu.timofte, vangool\}@vision.ee.ethz.ch} \\
	\texttt{\{liguanju, wujiqing9\}@gmail.com}
}

\iclrfinalcopy 
\begin{document}

	\maketitle
	
	\vspace{-2em}
	\begin{abstract}
		This paper introduces a divide-and-conquer inspired adversarial learning (DACAL) approach for photo enhancement. The key idea is to decompose the photo enhancement process into hierarchically multiple sub-problems, which can be better conquered from bottom to up. On the top level, we propose a perception-based division to learn additive and multiplicative components, required to translate a low-quality image or video into its high-quality counterpart. On the intermediate level, we use a frequency-based division with generative adversarial network (GAN) to weakly supervise the photo enhancement process. On the lower level, we design a dimension-based division that enables the GAN model to better approximates the distribution distance on multiple independent one-dimensional data to train the GAN model. While considering all three hierarchies, we develop multiscale and recurrent training approaches to optimize the image and video enhancement process in a weakly-supervised manner. Both quantitative and qualitative results clearly demonstrate that the proposed DACAL achieves the state-of-the-art performance for high-resolution image and video enhancement.
	\end{abstract}
	
	\section{Introduction}
	
	Despite many mobile camera technological advances we have today, our captured images often still come with limited dynamic range, undesirable color rendition, and unsatisfactory texture sharpness. Among many possible causes, low-light environments and under/over-exposed regions usually introduce severe lack of texture details and low-dynamic range coverage, respectively. Another critical issue is the amplification (during the enhancement process) of noise in the dark and/or texture-less regions, where the enhancement may not even be necessary. Due to these issues, images acquired under different conditions, or different parts of a single image, may require separate 
	enhancement operations. In this context, customarily used context/content agnostic enhancement methods often lead to a poor performance in the overall visual assessment.
	Therefore, comprehensive methods that improve the perceptual quality of images are in high demand.
	
	In recent years, deep image enhancement methods have  demonstrated their superiority for color enrichment, texture sharpening as well as contrast adjustment. A large amount of works like~(\cite{dong2014learning,kim2016accurate,mao2016image,shi2016real,wang2018fully,timofte2018ntire,nah2017deep,zhang2017beyond,zhang2016colorful,yu2018deepexposure,wu2018deep,yang2018image,wang2019under}) have made great success in one of such sub-tasks. However, in the real-world setting it is non-trivial to decompose the low-quality factors and treat them respectively from a low-quality images. Another family of approaches (e.g., \cite{gharbi2017deep, ignatov2017dslr,liu2018multiple,huang2018range,chen2018deep}) learns the enhancement as some non-linear mapping between low and high-quality image pairs, in a supervised manner. These methods require the desired high-quality images to be well aligned with the low-quality ones. Such image pairs are collected either by manually enhancing the low-quality images (using professional artists) or by using a calibrated rig of low and high-end cameras, with a very small baseline. Unfortunately, the data collection for these setups is not convenient, mainly due to the expensive manual retouching process and the physical limitation for small baseline. Moreover, fully-supervised methods often lack flexibility towards new domain adaptation, thus requiring low-high paired acquisitions for every low-end camera. 
	
	To address the limitations of fully-supervised methods, the weakly-supervised deep photo enhancement methods (e.g.,~\cite{ignatov2018wespe,chen2018deep}) have emerged recently. These methods merely require a set of target good quality images, whose contents and viewpoints share some similarity with those of the low-quality image domain, without requiring image acquisitions for the same scenes. 
	On the collected unpaired low- and high-quality images, such kind of methods typically apply generative adversarial networks (GANs), which alternately optimizes its two components (i.e., generator and discriminator) using a min-max objective. The discriminator is optimized to approximate the desired-quality image distribution, and the generator is guided by the discriminator to learn the enhancement map. While achieving some success, they still have drawbacks when enhancing mixed visual perceptions or treating high-resolution images. Without the implementation of a pixel-wise loss function in a supervised setup, it appears exceptionally difficult to simultaneously enhance multiple perceptual components (e.g., color, texture and illumination). Besides, due to the computational complexity and unstable adversarial training, the existing image enhancement methods treat high-resolution images using either the downscaled version or patch-wise processing strategies, neither of which are optimal. It is needless to mention that the downscaling loses images details, whereas, patch-wise enhancement does not maintain the spatial consistency over the whole image.

	\begin{figure*}[t]
		\begin{center}
			\includegraphics[width=0.9\linewidth]{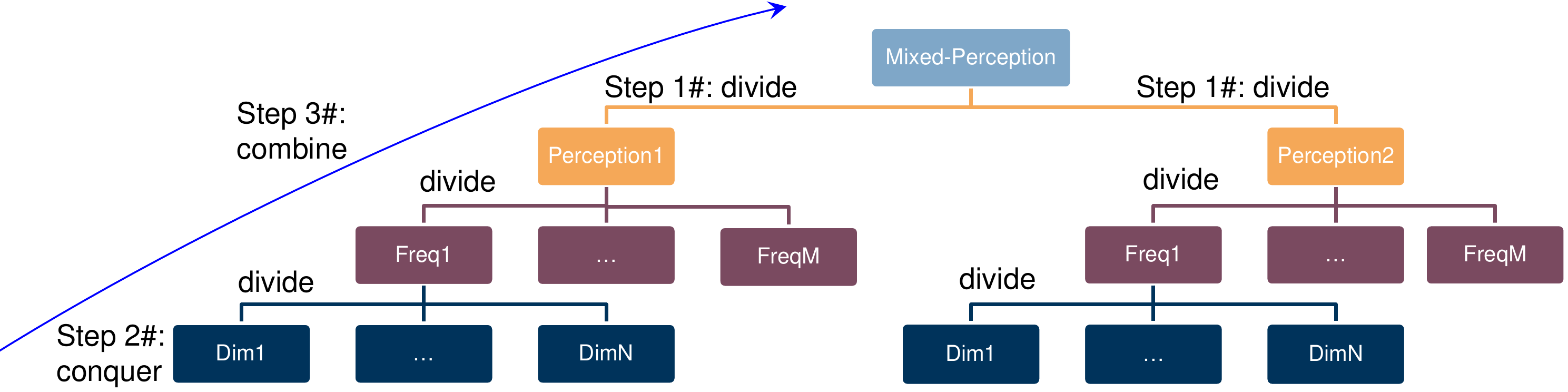}
		\end{center}
		\caption{Conceptual illustration of the proposed  adversarial learning approach of divide-and-conquer (DACAL). It decomposes the high-quality image distribution learning problem into perception-, frequency- and dimension-based sub-problems in a hierarchical structure. Following the division, the sub-problems are conquered and combined in a deep adversarial training fashion. }
		\label{fig:dacal}
	\end{figure*}
	
	In this paper, inspired by the concept of the divide-and-conquer (DAC) paradigm, we aim at decomposing the original problem of high-quality image distribution learning into sub-problems, which can be better addressed from bottom to up. For this purpose, we propose a DAC inspired adversarial learning approach\footnote{There exist some networks like (\cite{romaniuk1993divide,ghosh2017divide,nowak2018divide}) applying the paradigm of DAC to general neural networks. While we found two concurrent works (\cite{kim2019jsi,lin2019coco}) introduce the DAC concept to adversarial networks, they merely study the decomposition of generator. In contrast, our approach applies the DAC concept to the whole adversarial learning process.} to better deal with the high-resolution photo enhancement problem. As depicted in Figure~\ref{fig:dacal}, our basic learning process innovatively introduces the following three parts to the context of adversarial learning for high-resolution image and video enhancement.
	
	\begin{itemize}
		\item \textbf{Divide:} We hierarchically decompose the problem into three sub-problems: perception-based (additive and multiplicative maps), frequency-based (low and high frequencies), and dimension-based (multiple one dimensions) data distribution learning problems (from top to bottom). The third sub-problem that learns for each dimension separately is the basics for the original problem. In terms of implementation, these divisions are respectively carried out by:  (a) perception-wise branching of the generator; (b) frequency-wise division of the discriminator's inputs; and (c) dimension-wise division of the discriminator's output.

		\item \textbf{Conquer}: To conquer the base sub-problems, we exploit an adaptive sliced Wasserstein GAN (AdaSWGAN) model. The basic idea of the AdaSWGAN is to first factorize high-dimensional distributions into their multiple one-dimensional marginal distributions, followed by the  approximation of each one-dimensional distributions independently.

		\item \textbf{Combine}: Our model combines the solutions from bottom to top by, (a) aggregating the approximated one-dimensional distributions to obtain the frequency-based solutions; (b) averaging frequency-based solutions to obtain the perception-based solutions; (c) summing two perception-based solutions to learn the high-quality image distribution.
		
	\end{itemize}

	\section{Proposed Method}
	\label{proposed}
	
	\begin{figure*}[t]
		\begin{center}
			\includegraphics[width=1\linewidth]{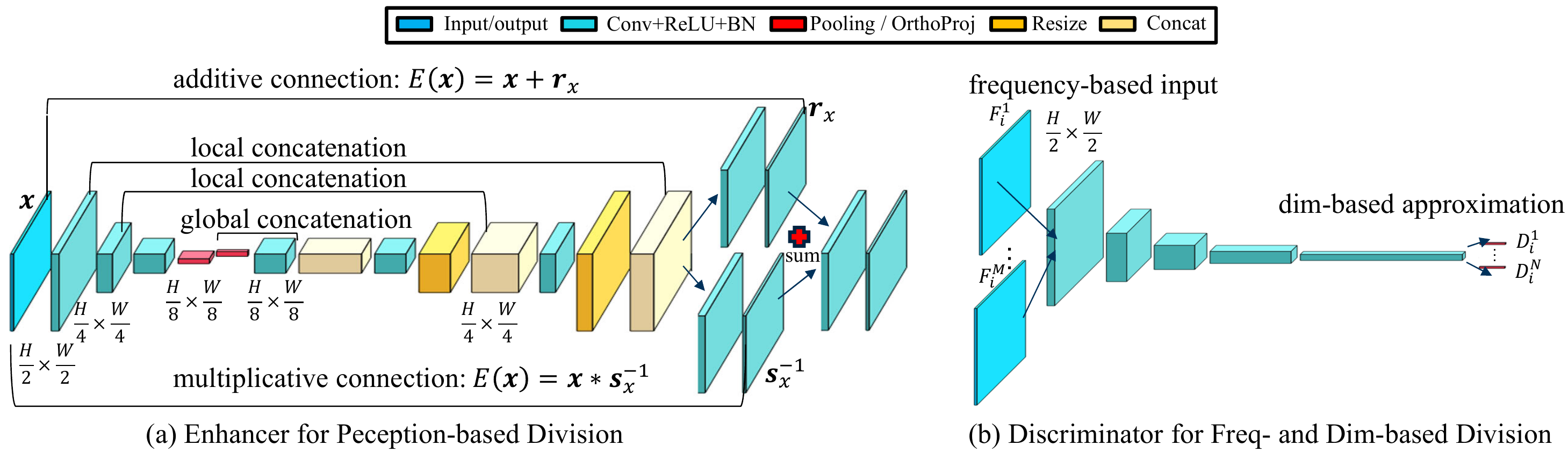}
		\end{center}
		\caption{Illustration of the proposed network design for our adversarial learning approach of divide-and-conquer. Specially, we apply the perception-based division to the generator, the frequency-based one to the discriminator input, and the dimension-based one to the discriminator output. }
		\label{fig:dacal_net}
	\end{figure*}

	\subsection{Perception-based Division}
	
	Given an input low-quality image $\bm{x} \in X$, image enhancement seeks a mapping function $E: X \rightarrow Y$, such that $\bm{y}=E(\bm{x})$, where $\bm{y}\in Y$ is the desired-quality image. We suggest to divide the mapping function learning problem into two subproblems, which learn additive and multiplicative perceptions respectively. On one hand, learning additive map is encouraged by the success of residual learning on image colorization and super-resolution {\it et cetera}. They apply the additive mapping $E(\bm{x})=\bm{x} + \bm{r}_x$ (where $\bm{r}_x$ is the residual component) to learn additive differences from low-quality images to desired-quality ones. 
	On the other hand, learning multiplicative map is inspired by existing works like (\cite{huang2009new, bioucas2010multiplicative,aubert2008variational, engel2016photometrically,wang2019under,guo2017lime,ghosh2017divide,fu2016weighted}). Such works  apply the multiplicative mapping $E(\bm{x})=\bm{x}*\bm{s}_x^{-1}$ (where $\bm{s}_x$ 
	is the multiplicative component) to address the problem of multiplicative noise removal or illumination improvement successfully. 
	Accordingly, we introduce the perception-wise division to the generator so that the benefits of both additive and multiplicative map learning approaches can be fully exploited. Formally, the final map of the generator is
	\begin{equation}
		E(\bm{x}) = \bm{x} + \bm{r}_x + \beta (\bm{x}*\bm{s}_x^{-1}),
		\label{Eq1}
	\end{equation}
	where $\beta$ plays the trade-off between such two perception-based components\footnote{Our experiments were conducted with $\beta =1$, which can be chosen differently for known sub-tasks priors.} and the operator `*' means element-wise multiplication.

	To approximate the additive component $\bm{r}_x$ and the multiplicative component $\bm{s}_x$, we design a two-stream U-Net~(\cite{ronneberger2015u}) for the generator (i.e., photo enhancer). As shown in Figure~\ref{fig:dacal_net}, we follow the standard U-Net design to apply the skip connection across the encoder and decoder \footnote{The encoder includes groups of convolution, scaled exponential linear units (\cite{klambauer2017self}) and batch normalization (\cite{ioffe2015batch}), while the decoder consists of convolution, resizing operation based layers.}. As skip connections concatenate local feature maps, they are good for reliable local feature extraction. In addition, we investigate that global information (e.g., scene category, object class or overall illumination) is useful for individual pixels to guide their local adjustment. Accordingly, we exploit a global concatenation operation to embed high-level information into decoded feature maps. While the encoder can transform input images into low-dimensional features, it does not always achieve a global feature (with a vector form) when the input images are of different sizes. To address this issue, we additionally apply an average pooling operation with a fully convolutional layer to extract the global feature vector. As a result, this design enables our model to enhance full-resolution images while it is trained on down-scaled images. On top of the decoder, we design two different branches to learn the additive and multiplicative components separately. During the early training, we optimize the additive and multiplicative mappings alternately. In the end of training, we aggregate the resulting approximations from these two branches, followed by a group of regular convolutional network layers. 
	
	The idea of our designed two-stream U-Net is related to~\cite{chen2018deep,wang2019under}. For the additive component learning, \cite{chen2018deep} introduces a global U-Net. Compared to our model, the global U-Net from~\cite{chen2018deep} can only accept images of a fixed low-resolution due to the hard design (a fixed filter for the fully-connected layer) for the global feature extraction. For example, when feeding $8\times 8 \times 128$ feature map into the global feature extraction, it has to use $8\times8$ filters to obtain the global feature $1 \times 1 \times 128$. For different image sizes, the resulting embedded feature size will be different, say $4 \times 8 \times 128$. In such cases, the global feature extraction will fail. For the multiplicative component learning, \cite{wang2019under} designs a bilateral-filtering based network, while our enhancer adopts regular convolutions and resizing operations.

	\subsection{Frequency-based Division}
	
	For the supervised photo enhancement problem that requires corresponding ground-truth high-quality images for guidance, the common reconstruction loss can be used to train the proposed deep enhancer. In contrast, the weakly-supervised photo enhancement task merely has a set of good quality images for reference, and thus for this task we need to learn the map from the low-quality image domain to high-quality image domain. One of the most promising approaches is cyclic GAN models~(\cite{zhu2017unpaired}). Hence, we choose to adopt the cyclic GAN objective to train our developed enhancer:
	\begin{equation}
		\begin{aligned}
			\min_{E, \hat{E}} \max_{C, \hat{C}}    \mathcal{L}_{\text{GAN}}(E, C) +  \mathcal{L}_{\text{GAN}}(\hat{E}, \hat{C}) + \gamma_1 \mathcal{L}_{\text{cyc}}(E, \hat{E})+\gamma_2 \mathcal{L}_{\text{id}}(E, \hat{E}),
		\end{aligned}
		\label{Eq2}
	\end{equation}
	where $E: X \rightarrow Y$ and $\hat{E}: Y \rightarrow X$ are used for forward and backward image translation, respectively. They share the same design using the proposed two-stream U-Net. Here, $C$ and $\hat{C}$ are the corresponding discriminators (or critics) which are used to guide the training of $E$ and $\hat{E}$, respectively. 
	The GAN loss $\mathcal{L}_{\text{GAN}}(E, C)$ and $\mathcal{L}_{\text{GAN}}(\hat{E}, \hat{C})$ are first suggested to use the adaptive Wasserstein GAN loss~(\cite{chen2018deep}) as,
	\begin{equation}
		\begin{aligned}
			\mathcal{L}_{\text{GAN}}^{C}(E, C) & =  \mathbb{E}_{\bm{y} \sim P_{y}} [C(\bm{y})]  -\mathbb{E}_{\bm{\tilde{y}} \sim P_E}[C(E(\bm{x}))]
			+ \lambda  \mathbb{E}_{\hat{y} \sim P_{\hat{y}}} [\max(0,\|\nabla_{\hat{y}} C(\hat{\bm{y}})\|_2-1)], \\
			\mathcal{L}_{\text{GAN}}^{E}(E, C)  & =
			\mathbb{E}_{\bm{x} \sim P_{x}} [C(E(\bm{x}))],
		\end{aligned}
		\label{Eq3}
	\end{equation}
	where $\hat{\bm{y}}$ are random samples following the distribution $P_{\hat{\bm{y}}}$, which is uniformly sampled along straight lines between pairs of points sampled from $P_{y}$ and $P_E$, and $\nabla_{\hat{y}} C(\hat{\bm{y}})$ is the gradient w.r.t $\hat{\bm{y}}$.
	
	Standard cyclic GAN models suggest to use the cyclic consistency loss and identity mapping loss. We follow (\cite{chen2018deep}) to define these two losses based on mean square error (MSE): 
	\begin{equation}
		\begin{aligned}
			\mathcal{L}_{\text{cyc}}(E, \hat{E}) &  =  \mathbb{E}_{\bm{x} \sim P_{x}} [\|\hat{E}(E(\bm{x})) -\bm{x}\|_2] +\mathbb{E}_{\bm{y} \sim P_{y}} [\|E(\hat{E}(\bm{y})) -\bm{y}\|_2],\\
			\mathcal{L}_{\text{id}}(E, \hat{E}) &  =  \mathbb{E}_{\bm{x} \sim P_{x}} [\|E(\bm{x}) -\bm{x}\|_2] +\mathbb{E}_{\bm{y} \sim P_{y}} [\|\hat{E}(\bm{y}) -\bm{y}\|_2].
		\end{aligned}
		\label{Eq4}
	\end{equation}
	
	To further decompose the perception-based distribution learning, we follow (\cite{ignatov2017dslr, ignatov2018wespe}) to apply frequency-based division to the discriminator $C$. Specifically, we adopt the separation of enhancing on low frequencies (e.g., main structures, major contents and colors), and optimizing on high frequencies (e.g., textures and small details). For this purpose, we replicate the discriminators to specialize on various frequencies of input images. Without loss of generality, we study two frequencies for discriminators: For low frequency data, we utilize the RGB image blurred by a Gaussian kernels, whereas, the grayscale component of the discriminator input images is used as the high frequency data. Both are equal in architecture and parameter values, and their GAN losses are averaged in the end. In this case, individual discriminators are encouraged to focus on an easier distribution approximation task, and thus become more reliable to supervise the training of the proposed photo enhancer.
	
	\begin{figure*}[t]
		\begin{center}
			\includegraphics[width=0.75\linewidth]{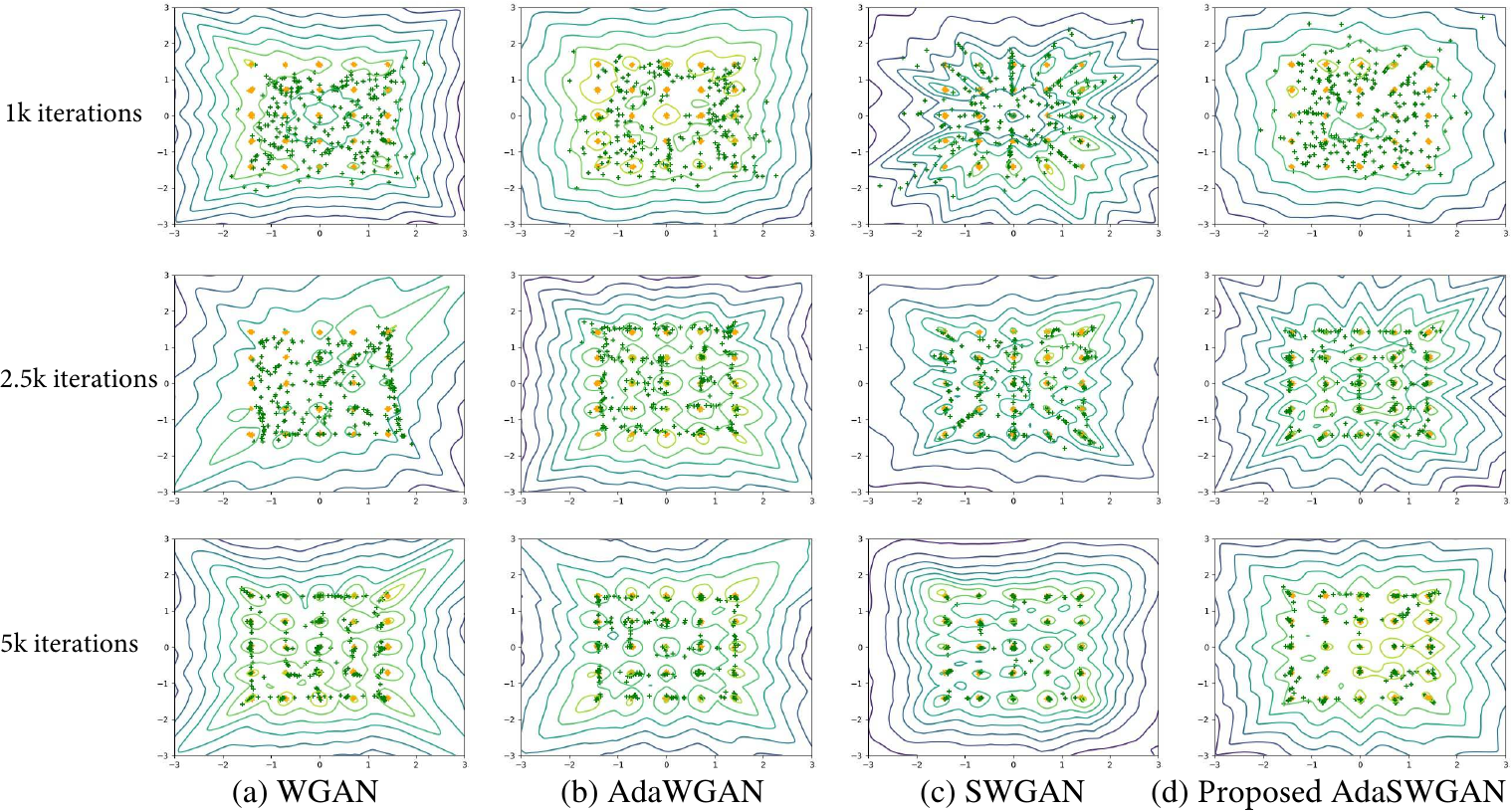}
		\end{center}
		\caption{Results of WGAN~(\cite{gulrajani2017improved}), AdaWGAN~(\cite{chen2018deep}), SWGAN~(\cite{wu2017sliced}) and the proposed AdaSWGAN (three rows indicate training for 1000, 2500 and 5000 iterations resp.) for the 25 Gaussians dataset. The orange points represent the target distribution, green points are the generated samples, and the curves indicate the value surfaces of the discriminators of compared models.}
		\label{fig:slicedDim}
	\end{figure*}
	
	\subsection{Dimension-based Division and Solution}
	
	While the frequency-based division for the discriminator's input can reduce the complexity of training Wasserstein GAN to some extent, it still suffers from the unstable training issue when the raw images lie in a high dimensional space, where we can only select samples very sparsely during training. To address this issue, we suggest to further decompose the challenging estimation of a high-dimensional distribution into simpler estimation of multiple one-dimensional distributions, which is the underlying idea of sliced Wasserstein distance computation. 
	To achieve this goal, we exploit an adaptive version of sliced Wasserstein GAN (AdaSWGAN) loss:
	\begin{equation}
		\begin{aligned}
			\min_{E} \max_{C} \; \int_{\theta \in \mathbb{S}^{n-1}} \Bigl(\mathbb{E}_{\bm{y} \sim P_y} [C(\bm{y})]  -\mathbb{E}_{\bm{\tilde{\bm{y}}} \sim P_E} [C(E(\bm{x}))] \Bigr) +  \lambda \mathbb{E}_{\hat{\bm{y}} \sim P_{\hat{\bm{y}}}}
			[\max(0, \|\nabla_{\hat{\bm{y}}} C(\hat{\bm{y}})\|_2 - 1)],
		\end{aligned}
		\label{Eq5}
	\end{equation}
	where $\theta$ is the orthogonal projection matrix for the decomposition from high-dimensional data to independent one-dimensional ones under $C$, $\hat{\bm{y}}$ denotes random samples following the distribution $P_{\hat{\bm{X}}}$ which is sampled uniformly along straight lines between pairs of points sampled from $P_X$ and $P_G$, and $\nabla_{\hat{y}} C(\hat{\bm{y}})$ is the gradient with respect to $\hat{\bm{y}}$. As the parameter $\lambda$ weights the gradient penalty, it can highly influence the trade-off between the original sliced Wasserstein distance and the Lipschitz constraint. To reach a good balance, we adopt an adaptive learning manner to update the hyperparameter by observing the moving average of gradients:
	\begin{equation}
		\begin{aligned}
			\overline{\nabla_{\hat{y}} C_(\hat{\bm{y}})} = \eta\overline{\nabla_{\hat{y}} C(\hat{\bm{y}})} +(1-\eta)\frac{\nabla_{\hat{y}} C(\hat{\bm{y}})}{\lambda},
		\end{aligned}
		\label{Eq6}
	\end{equation}
	where $\lambda$ is the weight for gradient penalty, and $\eta$ is the decay constant. When $\overline{\nabla_{\hat{y}} C(\hat{\bm{y}})}$ is larger than a fixed upper bound $\tau$, the relative influence of $\lambda$ becomes too small and therefore the penalty is not strong enough to satisfy the Lipschitz constraint. In this case, the weight $\lambda$ should be increased with a certain scale (in our experiments, we always set 2 times bigger). Otherwise, $\lambda$ is decreased to half of the current value.
	
	During training, the parameterized projection matrices $\theta$ should remain orthogonal. For this requirement we first initialize the parameters with random orthogonal matrices through QR decomposition, then update them on a Stifel manifold during training. Particularly, we follow~(\cite{huang2017riemannian,wu2017sliced}) to optimize the orthogonal matrices $\theta$ on Stiefel manifolds.

	Few variations of sliced Wasserstein GANs~(\cite{deshpande2018generative,Ishan2019sliced, wu2017sliced}) are proposed in the recent years. Differently from (\cite{deshpande2018generative,Ishan2019sliced}), we use the parameterized projection matrix to optimize the sliced Wasserstein distance approximation. When compared to (\cite{wu2017sliced}), we adopt an adaptive weight scheme to adjust the penalty weight. To show the advantage of the proposed AdaSWGAN, we follow the sampling strategy of (\cite{gulrajani2017improved}) to generate 25 Gaussian distributed data. On the toy dataset, we compare it against the four existing Wasserstein GAN models\footnote{For fair comparison, same generator and discriminator architectures (of \cite{gulrajani2017improved}) were used for all compared methods}. The results in Figure~\ref{fig:slicedDim} show that the proposed AdaSWGAN has a better convergence to fit the toy data than the other compared methods.

	\begin{figure*}[t]
		\begin{center}
			\includegraphics[width=1\linewidth]{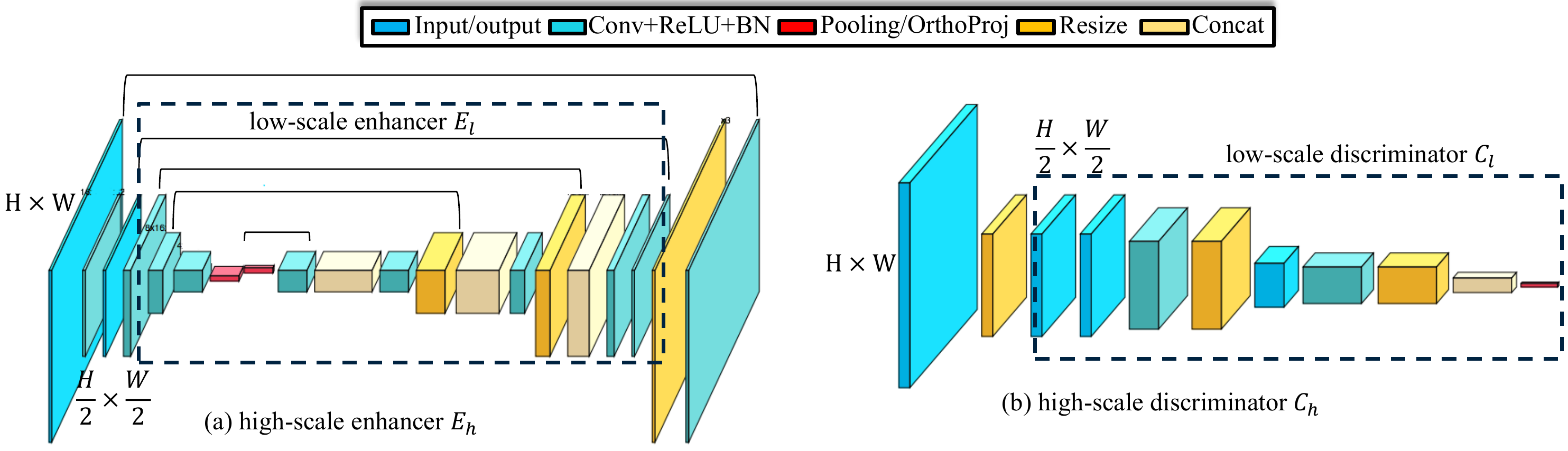}
		\end{center}
		\caption{Illustration of the multiscale network design for our adversarial learning approach of divide-and-conquer }
		\label{fig:multiscale_dacal}
	\end{figure*}

	\section{Multiscale Network Training}
	
	To improve the enhancement on high-resolution images, we introduce a multiscale enhancer that is trained on images with various resolutions. This design allows for a wider range of learned receptive fields, from coarse to fine. 
	The multiscale enhancer consists of various enhancers at multiple scales. Let's take a two-scale case for example. The two enhancers are denoted as $E_l, E_h$, which enhance images at resolution of $512 \times 256$ and $1024 \times 512$, respectively. The multiscaled design of enhancers is expected to effectively aggregate the visual features over various scales. 
	
	As shown in Figure \ref{fig:multiscale_dacal}, the high-scale enhancer $E_h$ has the same structure as the low-scale enhancer $E_l$. To transmit the low-scale information smoothly from the low-scale to the high-scale components, the high-scale enhancer also inherits the learned parameters from the low-scale enhancer. In addition, it repeats some more layers on the bottom and the top of the low-scale enhancer. Specially, it adds one more down-sampling component including convolution, scaled exponential linear units, and batch normalization at the bottom. One more up-sampling component is applied to the top. It contains an up-scale resizing and a convolution operation. Following the lower-scale enhancer design, we apply the same two-stream design to learn the additive and multiplicative components for the high-resolution image enhancement as well. 
	
	It is known that synthesizing high-resolution images yields significant challenges to the GAN training. For example, differentiating high-resolution real and generated images requires the discriminator to have a deeper architecture or convolutions with a larger receptive field. But such two solutions will often lead to overfitting and out of memory issues. Besides, due to the memory limit, we have to use smaller minibatches that will further harm training stability. 
	Therefore, as depicted in Figure \ref{fig:multiscale_dacal}, we also follow the design of multiscale enhancer to exploit multiscale discriminators for more reliable training guidance on different resolutions. Similarly, we insert some more layers into the bottom of the lower-scale discriminator to construct higher-scale discriminators for more reliable training guidance on higher-resolutions.
	
	For more details on the design of the multiscale enhancer and discriminator, please refer to the appendix. We train the proposed multiscale enhancer with three stages. At the first stage, we train the lower-enhancer $E_l$ using either reconstruction loss in the supervised case or using the proposed frequency- and dimension-based GAN loss for the weakly-supervised case. At the second stage, we tune the parameters of the newly added layers in the higher-scale enhancer $E_h$ while fixing the parameters of the lower-scale enhancer $E_l$. Finally, we train the multiscale enhancer as a whole.
	
	To overcome the unstable GAN training issue on high-resolution images, there exist various cascaded or multiscaled network designs like (\cite{denton2015deep,dosovitskiy2016generating,gatys2016image,johnson2016perceptual,zhang2017stackgan,huang2017stacked,karras2017progressive,wang2018high}) that synthesize high-resolution images. However, such works rarely study the multiscaled network based on the U-Net design, which has shown its advantages over other networks for image enhancement~(\cite{chen2018deep,huang2018range}). 
	In contrast, our multiscale enhancer is built upon a two-stream U-Net based model, which is tailored for better high-resolution image enhancement.  
	
	\begin{figure*}[t]
		\begin{center}
			\includegraphics[width=1\linewidth]{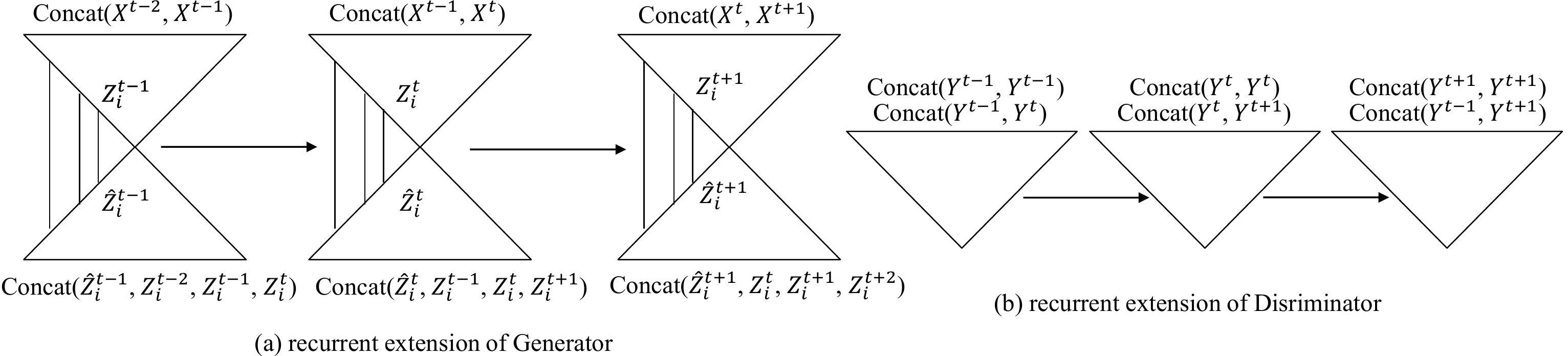}
		\end{center}
		\caption{Illustration of the recurrent network design for our adversarial learning approach of divide-and-conquer. Here $X$ and $Y$ denote the original inputs for generator and discriminator respectively, and $Z$ and $\hat{Z}$ represent the original encoded and decoded features from generator. The recurrent generator accepts the concatenation of the consecutive frames for input, and propagates the feature maps from the previous and next frames by extending the original skip connection. The recurrent discriminator receives the original inputs as well as the concatenation of continuous frames to simultaneously learn the individual and sequential distributions. }
		\label{fig:recurrent_dacal}
	\end{figure*}
	
	\section{Recurrent Network Training}
	
	Considering the balance between temporal smoothing performance and efficiency, we suggest to extend the designed network with recurrent approaches for video enhancement. For the recurrent extension of the generator that is based on the U-Net design, we propagate the input data as well as the encoder features over time. The idea is illustrated in Figure \ref{fig:recurrent_dacal}. More specially, we concatenate the adjacent two frames for low-level information based propagation. In this case, the temporal correlation of continuous frames is able to offer more reliable guidance for the update on the encoder parameters. Comparing with the original U-Net's decoder that accepts the encoder feature using skip connection, we propose to stack the corresponding feature maps from the previous and next frames as well. This is expected to propagate high-level semantics along the time dimension. For the discriminator design, except feeding it with the single frames, we also input the concatenation of consecutive frames for sequential distribution learning. For more stable learning on temporal information, we also apply bidirectional training strategy, which alternately uses the forward and backward sequential frames for training.
	
	While there are some classical recurrent networks/units like Long Short-Term Memory (LSTM) \cite{hochreiter1997long} and Gated Recurrent Unit (GRU) \cite{chung2014empirical, cho2014properties}, we are more favor of cheaper recurrent unit to improve the temporal consistency for video enhancement. Our recurrent unit design is inspired by  \cite{fuoli2019efficient} which introduces high-dimensional latent states to propagate temporal information between frames in an implicit manner in the context of shuffling network.
	
	\section{Experiment}
	
	\subsection{Supervised Image Enhancement}
	
	This task aims at learning the enhancement mapping between low- and high-quality image pairs in a supervised fashion. We use the commonly-used MIT-Adobe FiveK~(\cite{bychkovsky2011learning}) dataset for this task. The dataset consists of 5,000 high-resolution (higher than 1K) images, each of which is retouched by five experts. For the evaluation, we use 2,250 images and their retouched versions from photographer C for training, and the remaining 498 images are used for testing. Distinct from most of existing works, we go for high-resolution image enhancement. 
	To study this task, we compare our proposed adersarial learning approach of divide-and-conquer (DACAL) against state-of-the-art methods including White-Box (WB)~(\cite{hu2018exposure}), Distort-and-Recover (DR)~(\cite{park2018distort}), DSLR-Photo Enhancement (DPED)~(\cite{ignatov2017dslr}), and the supervised version of the Deep Photo Enhancement (DPE) method~(\cite{chen2018deep}). Except DPED, the rest three methods are originally designed for down-scaled image enhancement. In contrast, our model can be trained on multiscaled images. For details on our training details\footnote{The code will be released once the paper is published.}, please refer to the appendix. As the dataset has paired original/retouched images, we use both the Peak Signal-to-Noise Ratio (PSNR) and the multi-scale structural similarity index measure (SSIM)\footnote{For short, we denote the mutli-scale SSIM as SSIM throughout the paper.}~(\cite{wang2003multiscale}) to evaluate the comparing methods.

	\begin{table}[t]
		\centering
		\footnotesize
		\caption{PSNR and SSIM results for the MIT-Adobe FiveK~(\cite{bychkovsky2011learning}) test images. Here, WB and DR indicate the White-Box and Distort-and-Recover methods, respectively. DACAL$_{l_1}$, DACAL$_{l_2}$, DACAL$_{l_3}$ and DACAL$_{l}$ represent the use of individual additive, individual multiplicative, multiplicative cascaded by additive, and our suggested parallel fusion (two-stream strategy), respectively. DACAL$_h$ is our higher-scale version. PSNR$_d$/SSIM$_d$ and PSNR$_f$/SSIM$_f$ indicate the results on downscaled images and full-resolution images, respectively.}
		\begin{tabular}{cccccccccc}
			\toprule
			& WB & DR & DPED & DPE & DACAL$_{l_1}$ & DACAL$_{l_2}$ & DACAL$_{l_3}$  & DACAL$_l$ & DACAL$_h$ \\ \hline
			PSNR$_d$ & 18.86   & 21.64 & 21.05  & 22.10 & 22.73 & 22.99 & 23.01  &  23.52 & \textbf{24.15} \\ 
			PSNR$_f$ & 19.09   & 21.52 & 20.86  & 21.65	& 22.43  & 22.69   & 23.02 & 23.56 & \textbf{24.07} \\
			\bottomrule
			SSIM$_d$  & 0.928  & 0.936 & 0.922 & 0.947 & 0.958 & 0.942  & 0.949 & 0.959 & \textbf{0.962}  \\ 
			SSIM$_f$ & 0.920   & 0.922 &  0.916 & 0.894 & 0.948 & 0.942 & 0.940  & 0.954 & \textbf{0.956} \\ 
			\bottomrule
		\end{tabular}
		\label{tab:quanti_adobe}
		\vspace{-0.2cm}
	\end{table}

	\begin{figure*}[t]
		\begin{center}
			\includegraphics[width=1\linewidth]{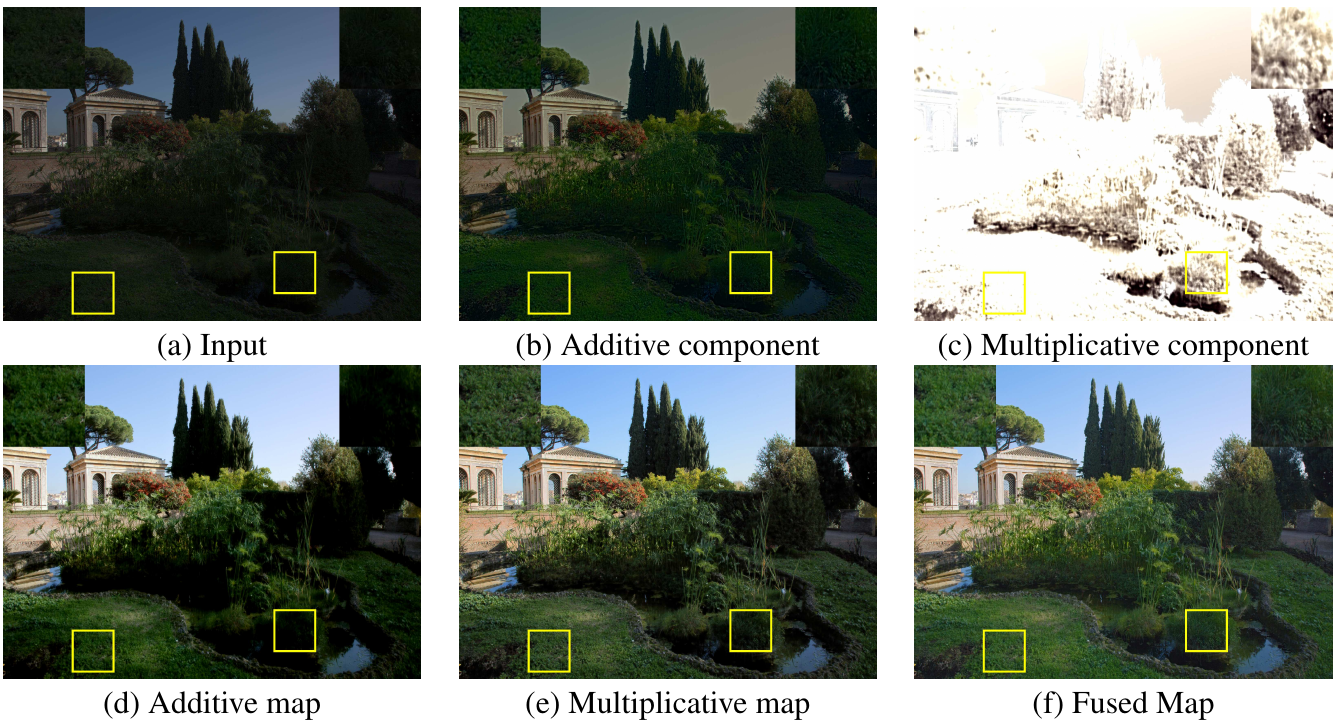}
		\end{center}
		\caption{Visual results of additive, multiplicative components and their combination. It is best to
			zoom in on the high-resolution pictures.}
		\label{fig:add_multi}
	\end{figure*}
	
	Table~\ref{tab:quanti_adobe} summarizes the PSNR and SSIM results on both downscaled images of size $512 \times 256$ and full-resolution ones. Compared to the state-of-the-art methods, our proposed DACAL achieves the highest PSNR and SSIM. Note that our DACAL achieves the highest PSNR~\footnote{Recently the CVPR19 paper (\cite{wang2019under}) has corrected their model's PSNR from 30.80 dB to 23.04 dB, on its GitHub webpage.} (\textbf{24.15 dB}) on the MIT-Adobe FiveK dataset. In particular, we discover the PSNR and SSIM values of some competing methods like DPE decline by a certain margin when enhancing full-resolution images. In contrast, our proposed DACAL has relatively stable PSNR and SSIM values in the high-res case. For a quantitative perceptual quality comparison, we employ 20 Amazon Mechanical Turk (MTurk) workers to compare 100 image pairs, which are from our DACAL and the competing methods respectively. As shown in Table~\ref{tab:user_study}, the highest preference score shows the superiority of our DACAL.

	In addition, we conduct the ablation study on the proposed two-stream U-Net. Specially, we evaluate the separative use of additive and multiplicative streams denoted by DACAL$_{l_1}$ and DACAL$_{l_2}$, respectively. As shown in Table \ref{tab:quanti_adobe}, they are outperformed by our two-stream version DACAL$_{l}$.
	We also evaluate the performance by cascading these two streams: additive cascaded by multiplicative stream, or multiplicative cascaded by additive stream. As we find the former case works better than the later one in terms of PSNR and SSIM. Here, we only report the results of the former one DACAL$_{l_3}$. The improvement of our suggested two-stream version further justifies its effectiveness. In Figure \ref{fig:add_multi}, we also study the visual results of the two components and their resulting maps. The results reflect that the additive one targets for better color and texture, and the multiplicative one targets on enhancing illumination. The result of final map shows the benefit of fusing additive and multiplicative streams.

	\subsection{Weakly-Supervised Image Enhancement}
	
	This task performs photo enhancement using one set of good-quality images instead of paired before/after images. For this task, we use the DPED dataset~(\cite{ignatov2017dslr}) and our collected HuaweiM20Pro-Flickr dataset. The DPED dataset is composed of images from smartphone cameras (i.e., iPhone 3GS, BlackBerry Passport and Sony Xperia Z) paired with images of the same scenescaptured by a DSLR camera (i.e., Canon 70D). We use the iPhone3GS-DSLR data (5,614 iPhone images of size $2048 \times 1536$, 5,902 DSLR images of size $3648 \times 2432$) to evaluate the competing methods on the full-resolution images in a weakly-supervised manner. As the dataset also contains 4,353 well-aligned $100 \times 100$ iPhone3GS-DSLR image patch pairs, we use its testing image patches to compute the PSNR and SSIM values for quantitative evaluation. As for our HuaweiM20Pro-Flickr dataset, we collect 5,138 images of size $1920 \times 1080$ by a Huawei Mate20 pro and use the downloaded 604 HDR Flickr images with around 1K resolution from~(\cite{chen2018deep}) for the target.
	
\begin{figure*}[t]
		\begin{center}
			\includegraphics[width=1\linewidth]{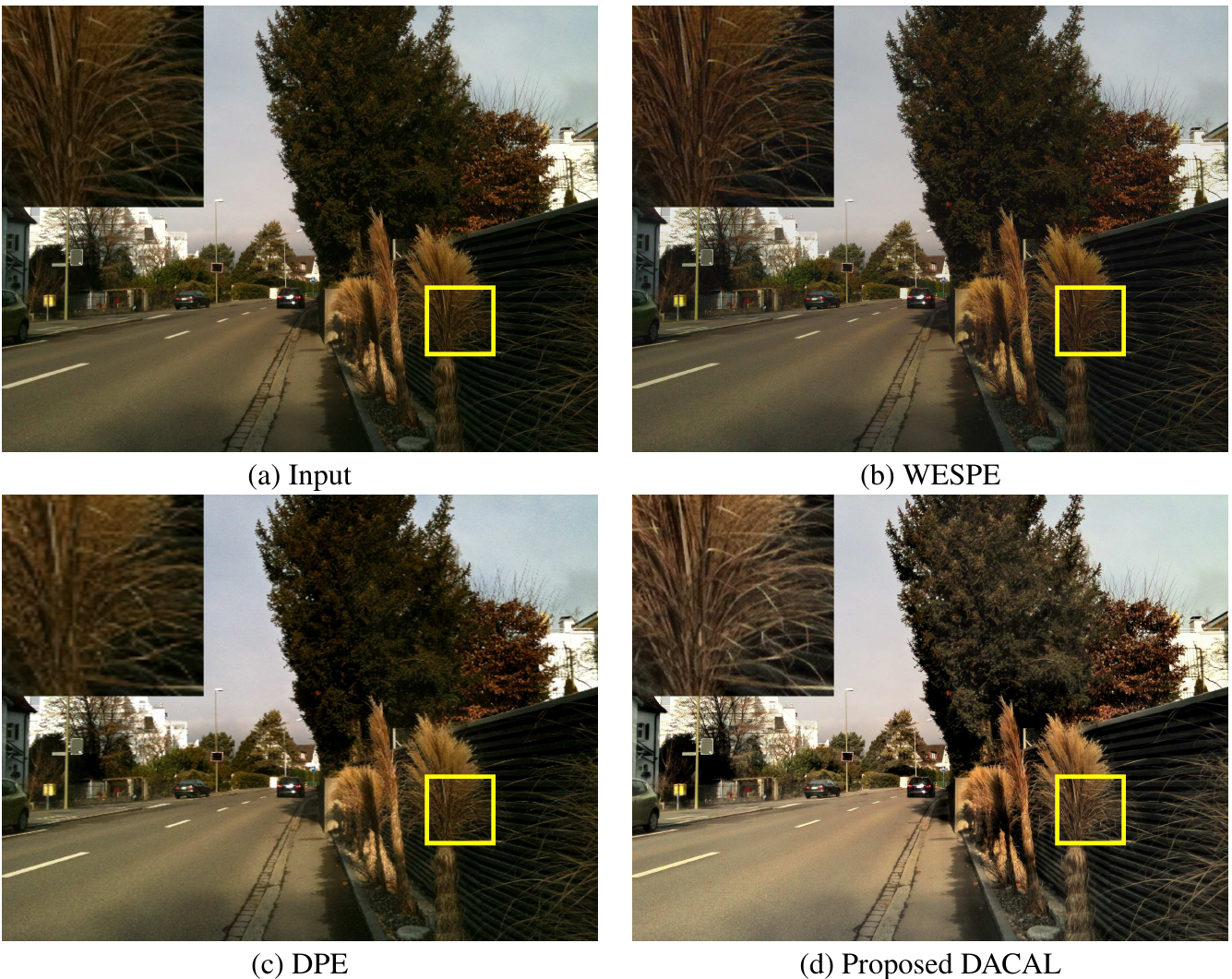}
		\end{center}
		\caption{High-resolution image enhancement results for the DPED iPhone3GS-DSLR data. It is best to zoom in on the high-resolution pictures.}
		\label{fig:enhanced_dped}
	\end{figure*}

\begin{figure*}[t]
		\begin{center}
			\includegraphics[width=1\linewidth]{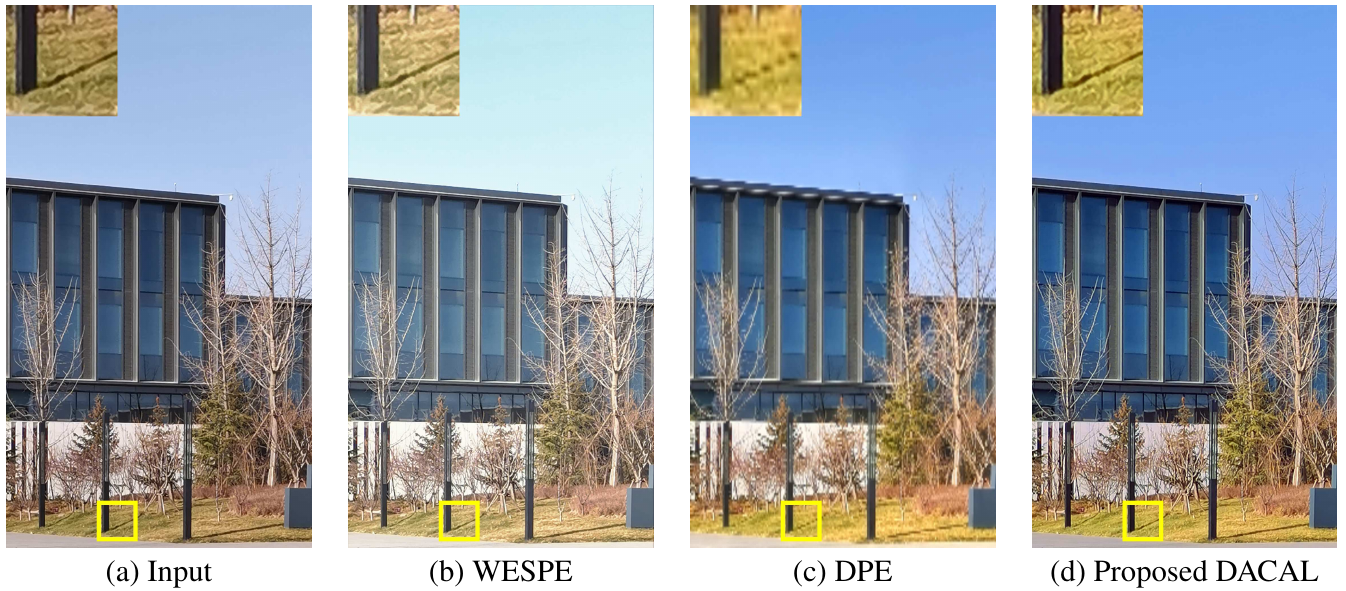}
		\end{center}
		\caption{High-resolution image enhancement results for our HuaweiM20Pro-Flickr data. It is best to zoom in on the high-resolution pictures.}
		\label{fig:enhanced_huawei}
	\end{figure*}

	Table~\ref{tab:quanti_dped} reports the PSNR and SSIM results of the evaluated methods on the DPED dataset. When compared with the state-of-art weakly-supervised methods WESPE~(\cite{ignatov2018wespe}) and DPE~(\cite{chen2018deep}), our DACAL reaches the highest PSNR (\textbf{20.90 dB}) and SSIM. The table also shows the improvement of the introduced frequency based GAN and sliced SWGAN loss. For a quantitative perceptual quality comparison on our DACAL, WESPE and DPE, we ask 20 MTurk works for user study on 120 image pairs for DPED and HuaweiM20Pro-Flickr. The results in Table~\ref{tab:user_study} show that our method achieves much higher preference score than all competing methods in terms of the sampled human perception. The qualitative results from Figure~\ref{fig:enhanced_dped} reflect that our DACAL has the best performance in terms of color rendering, texture sharpening, and illumination adjustment for the DPED iPhone images. In Figure~\ref{fig:enhanced_huawei}, we present enhancement results for our HuaweiM20Pro-Flickr dataset. We can find that WESPE tends to make the image over-exposed, while DPE has serious texture issues due to its design for low-resolution image enhancement. By comparison, our DACAL performs most promisingly for color, texture, and light condition improvements.

	\begin{table}[t]
		\centering
		\footnotesize
		\caption{PSNR and SSIM results for the DPED~(\cite{ignatov2017dslr}) test $100 \times 100$ image patches. Here, $l, f, d$ for DACAL represent the use of our proposed perception-based, frequency-based and dimension-based division respectively. DACAL$_h$ is our higher-scale version. }
		\begin{tabular}{ccccccc}
			\toprule
			& WESPE & DPE & DACAL$_{l}$ & DACAL$_{l+f}$ & DACAL$_{l+f+d}$ & DACAL$_h$ \\ \hline
			PSNR$_{100}$   & 17.45 & 18.53 & 19.62 & 20.01  & 20.43 & \textbf{20.90} \\ 
			SSIM$_{100}$   & 0.854 & 0.861 & 0.868 & 0.869  & 0.872 & \textbf{0.874}     \\ 
			\bottomrule
		\end{tabular}
		\label{tab:quanti_dped}
		\vspace{-0.2cm}
	\end{table}

	\begin{table}[t]
		\centering
		\footnotesize
		\caption{Preference ratio (one vote is lost for DACAL vs. Input) from MTurk user study }
		\begin{tabular}{ccccccc}
			\toprule
			DACAL vs. &Input  & White-Box & Distort-and-Recover & DPED & WESPE & DPE   \\ \hline
			Adobe & 395:4  &  389:11  & 303:97 & 365:35 & N/A & 370:30  \\ 
			DPED   & 384:16 & N/A & N/A & N/A & 384:16 & 344:56  \\ 
			HuaweiM20Pro-Flickr & 366:34   & N/A  & N/A & N/A & 341:59 & 360:40 \\ 
			\bottomrule
		\end{tabular}
		\label{tab:user_study}
		\vspace{-0.2cm}
	\end{table}

	\begin{figure*}[t]
		\begin{center}
			\includegraphics[width=0.92\linewidth]{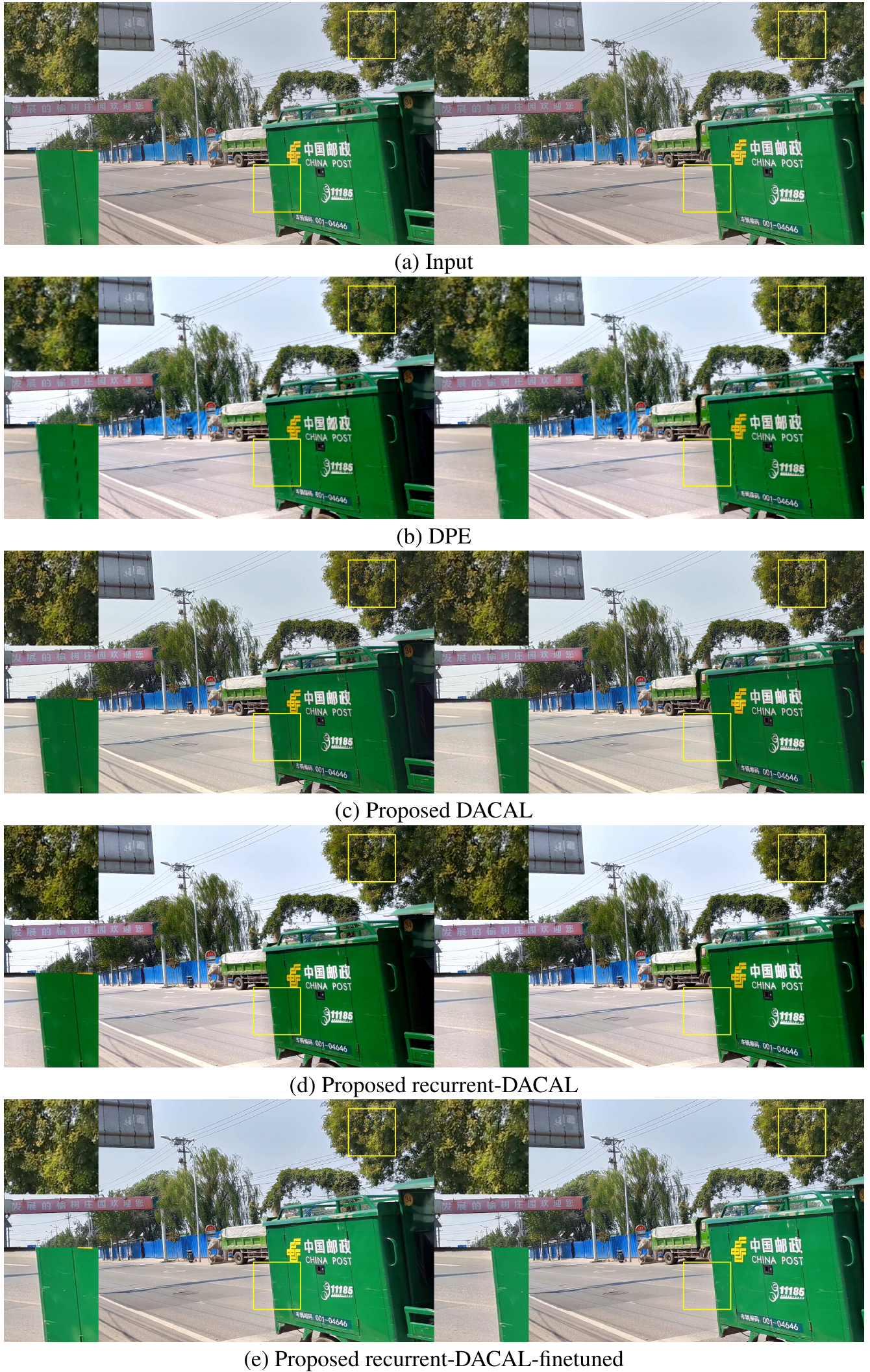}
		\end{center}
		\vspace{-1em}
		\caption{High-resolution video enhancement results  on two consecutive frames for the HuaweiP30Pro-PanasonicGH5S data. It is best to zoom in on the high-resolution pictures.}
		
		\label{fig:enhanced_huawei_video}
		\vspace{-1em}
		
	\end{figure*}

	\begin{figure*}[t]
		\begin{center}
			\includegraphics[width=0.92\linewidth]{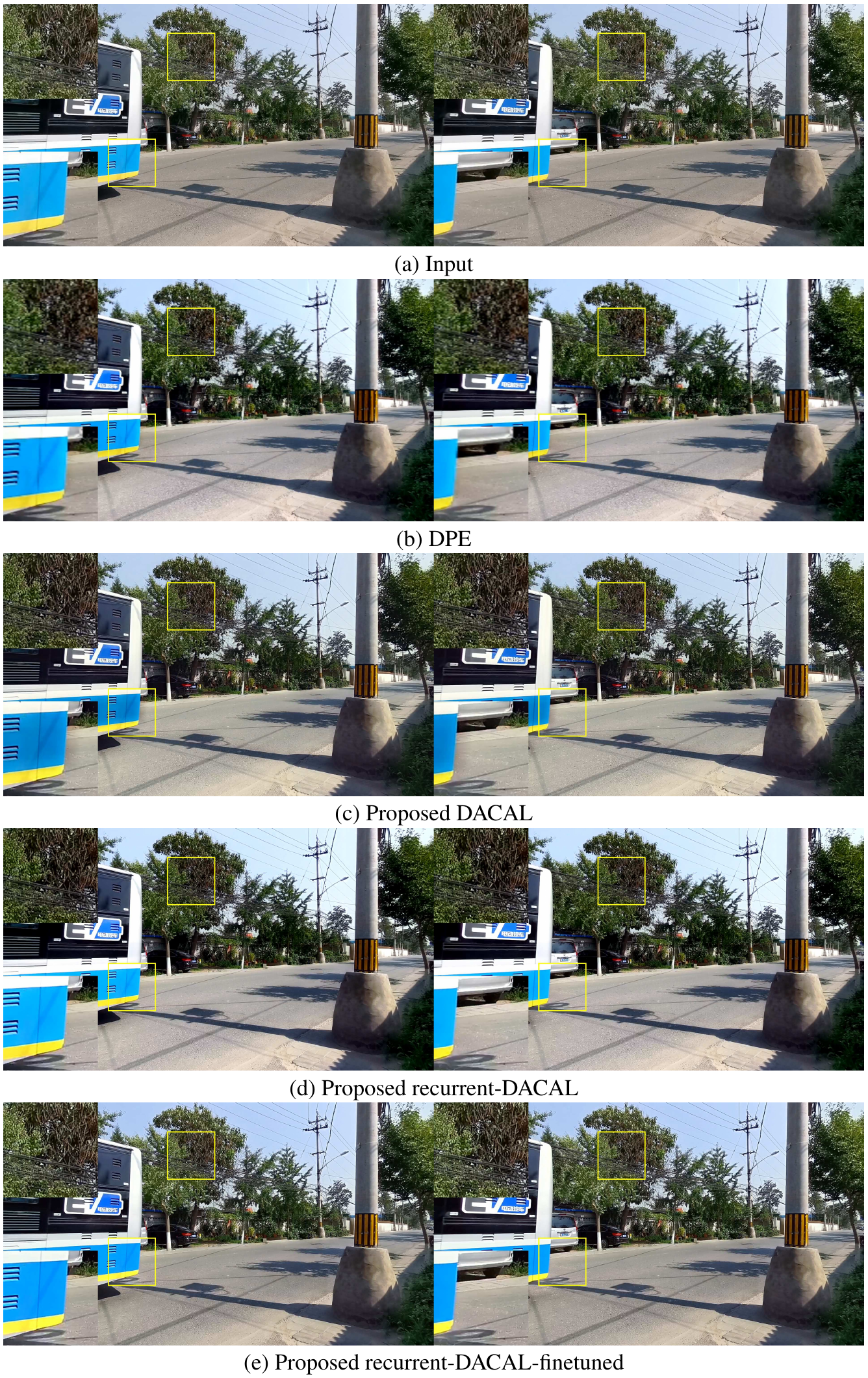}
		\end{center}
		\vspace{-1em}
		\caption{High-resolution video enhancement results  on two consecutive frames for the HuaweiP30Pro-PanasonicGH5S data. It is best to zoom in on the high-resolution pictures.}
		\label{fig:enhanced_huawei_video2}
		\vspace{-1em}
		
	\end{figure*}
	
	\subsection{Weakly-Supervised Video Enhancement}
	
	The goal of video enhancement is to enhance the perceived quality of individual frames without undermining their temporal consistency. To study this task, we collect a video dataset containing 1370 video pairs, which are captured by a Huawei P30 Pro and a Panasonic GH5s respectively. The length of each recording is between 5 and 10 seconds. Videos are captured by the P30 Pro in 29.95 FPS, while those taken by the Panasonic GH5s are of 25 FPS. The resolution of videos are mostly of $1920 \times 1080$. On this dataset, we use 1270 video pairs for training and the remaining 100 pairs for testing. While each video pair is captured at approximately the same location, it is still very challenging to align them well in both spatial and temporal domains. On the other side, there are very few video enhancement methods which are trained on such weakly-paired data. Accordingly, we can only evaluate the existing weakly-supervised per-frame enhancement methods like DPE. For comparison, we evaluate our proposed DACAL and the recurrent variant of DACAL as well. Besides, we also fine-tuned the recurrent DACAL on an image dataset that consists of 786 DSLR-quality and retouched images.
    
     Figure \ref{fig:enhanced_huawei_video} and Figure \ref{fig:enhanced_huawei_video2} show the enhancement results on two consecutive frames from two videos. Similar to the image enhancement case, the DPE method has the same issue on textures. In addition, we find there are some temporal incoherence issues for the per-frame enhancement methods like DPE and our proposed DACAL. For example, the moving objects often have some shadows or artifacts around the boundaries. In contrast, the recurrent extension of DACAL is able to alleviate the issue by using the temporal propagation strategy. Once the model is further fine-tuned on higher-quality reference images, the moving artifacts can be almost removed while improving the perceptual quality in terms of colorization, contrast and texture.

	\section{Conclusion}
	
This paper addresses mix-perception enhancement of images and videos with a divide-and-conquer inspired adversarial learning approach, which divides and merges perception-based, frequency-based and dimension-based sub-problems. For high-resolution image and video enhancement, we further suggest multiscaled and recurrent extensions of the basic model. Evaluations on both supervised and weakly-supervised image/video enhancement tasks demonstrate the clear superiority of our proposed enhancer against the competing methods.


	\appendix
	\section{Appendix}
	
	\begin{figure*}[!htbp]
		\begin{center}
			\includegraphics[width=1\linewidth]{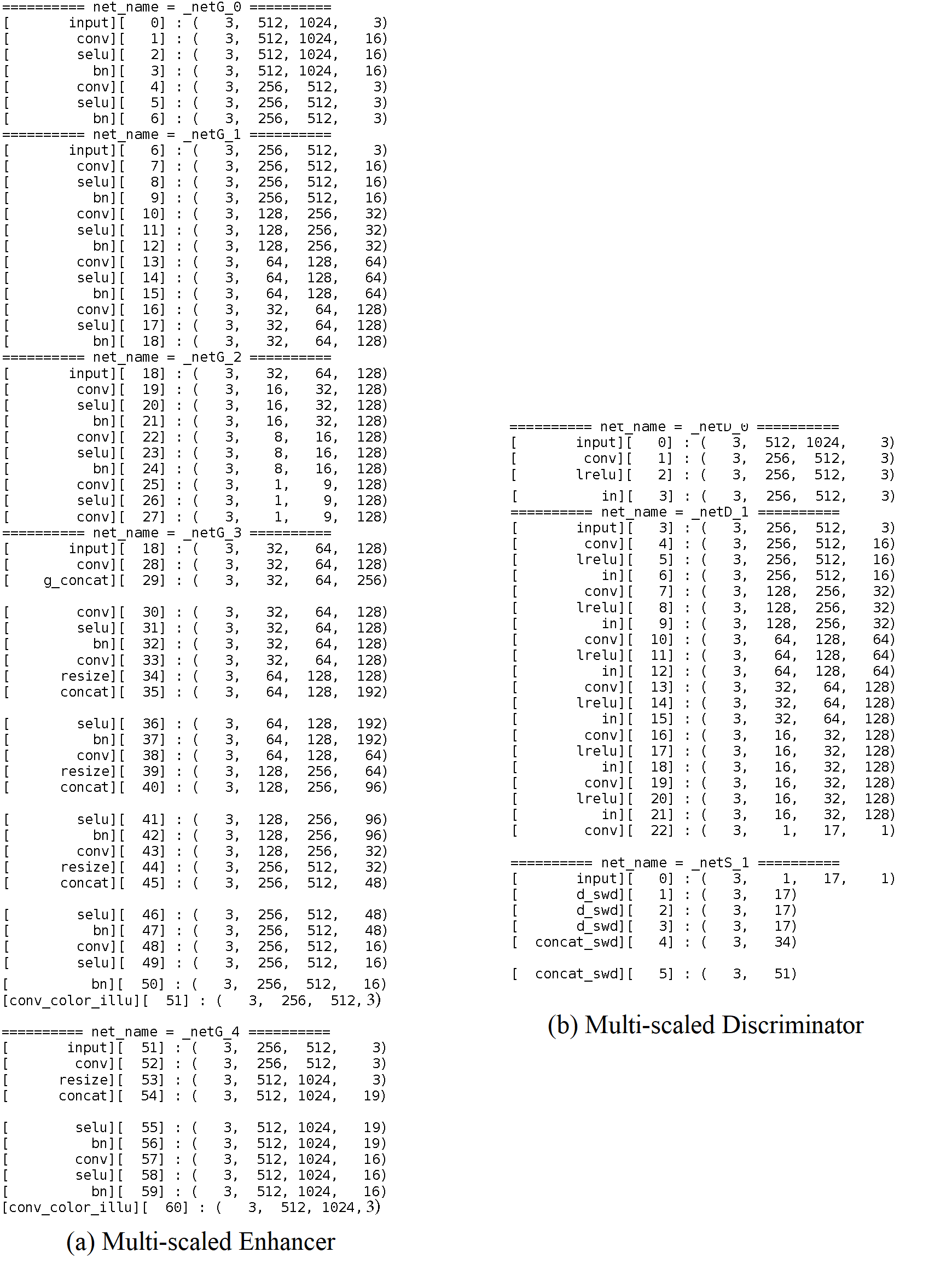}
		\end{center}
		\caption{Illustration of the used multi-scaled enhancer (a) and multi-scaled discriminator (b) for our proposed method}
		\label{fig:network}
	\end{figure*}
	
	\subsection{Network Architecture}
	
	In the paper, we use two-scale design for both enhancer and discriminator.
	Figure~\ref{fig:network} illustrates the network architecture of our proposed method. Specially, Figure~\ref{fig:network} (a) shows the network design of our multiscaled enhancer, consisting of 5 components: netG$_0$, netG$_1$, netG$_2$, netG$_3$ and netG$_4$. Among them, netG$_1$, netG$_2$, netG$_3$ form the lower-scale enhancer. Therefore, the higher-scale enhancer inherits netG$_1$, netG$_2$, netG$_3$ from lower-scale enhancer, and introduces two new components that are netG$_0$ and netG$_4$. Note that the layer named `conv-color-illu' realizes the two-stream branches, which contains separate groups of convolutions operations, and finally aggregates them using the Eqn.1 in the major paper. Figure~\ref{fig:network} (b) shows the structure of our multiscaled discriminator that contains netD$_0$, netD$_1$ and netS$_1$. The component netS$_1$ is designed for the computation of sliced Wasserstein distance. Analogously, netD$_1$ and netS$_1$ are from lower-scale discriminator. The higher-scale discriminators add one more component netD$_0$.

	\subsection{Training Details}
	
	For the supervised image enhancement task, we train the proposed two-scaled enhancer on one single NVIDIA TITAN Xp GPU with 12GB GPU memory. We first train the lower-scaled enhancer over down-scaled ($512 \times 256)$ images for 20 epochs, and then train higher-scaled enhancer through tuning the new parameters of netG$_0$ and netG$_4$ while fixing the parameters of netG$_1$, netG$_2$, netG$_3$  over higher-resolution ($1024 \times 512$) images for 20 epochs. Finally, we train the whole multi-scale enhancer for 20 epochs.
	For the weakly supervised image enhancement task, we train the two-scaled enhancer on two NVIDIA TITAN Xp GPUs, each of which has 12GB GPU memory. Similar to the supervised case, we use the same style to train the multi-scaled enhancer. For the training, we empirically set the hyperparameters $\lambda=10, \eta=0.99, \tau=0.05, \gamma_1 = 10000, \gamma_2 = 1000$ from the proposed GAN model. Note that it is still feasible for us to train our enhancer on higher-resolution images if we have more computational resources.

\end{document}

%% file: math_commands.tex
\usepackage{amsmath,amsfonts,bm}

\def\eqref#1{equation~\ref{#1}}

\def\1{\bm{1}}

\DeclareMathAlphabet{\mathsfit}{\encodingdefault}{\sfdefault}{m}{sl}
\SetMathAlphabet{\mathsfit}{bold}{\encodingdefault}{\sfdefault}{bx}{n}

